# *Science* Manuscript Template

## Title: Unfolding the Laws of Star Formation: The Density Distribution of Molecular Clouds

**Authors:** Jouni Kainulainen[1]*, Christoph Federrath[2], Thomas Henning[1]

**Affiliations:**

[1]Max-Planck-Institute for Astronomy, Königstuhl 17, 69117 Heidelberg, Germany.

[2]Monash Centre for Astrophysics, School of Mathematical Sciences, Monash University, Vic 3800, Australia.

*Correspondence to: jtkainul@mpia.de .

**Abstract**: The formation of stars shapes the structure and evolution of entire galaxies. The rate and efficiency of this process are affected substantially by the density structure of the individual molecular clouds in which stars form. The most fundamental measure of this structure is the probability density function of volume densities ($\rho$-PDF), which determines the star formation rates predicted with analytical models. This function has remained unconstrained by observations. We have developed an approach to quantify $\rho$-PDFs and establish their relation to star formation. The $\rho$-PDFs instigate a density threshold of star formation and allow us to quantify the star formation efficiency above it. The $\rho$-PDFs provide new constraints for star formation theories and correctly predict several key properties of the star-forming interstellar medium.

**Main Text:** The formation of stars is an indivisible component of our current picture of galaxy evolution. It also represents the first step in defining where new planetary systems can form. The physics of how the interstellar medium is converted into stars is strongly affected by the density structure of individual molecular clouds (*1*). This structure directly affects the star formation rates (*SFR*) and efficiencies (*SFE*) predicted by analytic models (*2-5*). Inferring this structure observationally is challenging, because observations only probe projected column densities. Hence, the key parameters of star formation models remain unconstrained. Here we present a technique that allows us to quantify the grounding measure of the molecular cloud density structure, the probability density function of their volume density ($\rho$-PDF).

The *SFRs* of molecular clouds are estimated in analytic theories from the amount of gas in the clouds above a critical density, $\rho_{crit}$ (*2-5*):

$$SFR = \frac{\varepsilon_{core}}{\phi} \int_{s_{crit}}^{\infty} \frac{t_{ff}(\rho_0)}{t_{ff}(\rho)} \frac{\rho}{\rho_0} p(s)ds \qquad (1)$$

where $s = \ln(\rho/\rho_0)$ is the logarithmic, mean-normalized density, and $s_{crit} = \ln(\rho_{crit}/\rho_0)$. We use the number density, $n = \rho/\bar{\mu}m_p$, where $\bar{\mu}$ is the mean molecular mass and $m_p$ the proton mass, as the measure of density. The parameter $\varepsilon_{core}$ in Eq. (1) is the core-to-star efficiency, giving the fraction of gas above $s_{crit}$ that collapses into a star. The $t_{ff}(\rho)$ is the free-fall time of pressure-less gas that approximates the star-formation timescale, and $\phi$ is the ratio of the free-fall time to the actual star formation timescale. The critical density, commonly referred to as the (volume) density threshold of star formation, indicates that stars form only above that density. Generally, the critical density depends on gas properties (*2-5*), but theoretical considerations suggest that it could be relatively constant under typical molecular cloud conditions (*5*).

The decisive density structure of molecular clouds is encapsulated in the function $p(s)$ describing the probability of a volume $dV$ to have a log density between $[s, s+ds]$, i.e., the $\rho$-PDF. In current understanding, the $\rho$-PDF is determined by supersonic turbulence that induces a log-normal $\rho$-PDF (*6-9*):

$$p(s) = \frac{1}{\sigma_s \sqrt{2\pi}} e^{-\frac{(s-\mu)^2}{2\sigma_s^2}}, \qquad (2)$$

where $\mu$ and $\sigma_s$ are the mean and width, respectively. The $\rho$-PDF width is linked to the turbulent gas properties through $\sigma_s^2 = \ln(1 + b^2 M_s^2 \frac{\beta}{\beta+1})$ (*10*), where $M_s$ (sonic Mach number) is a measure of the turbulence energy, $b$ is a parameter related to the turbulence driving mechanism (*9*), and $\beta$ is the ratio of thermal to magnetic pressures.

Despite their decisive role for star formation, the $\rho$-PDF function and the critical density are not well constrained observationally. Instead, studies have measured their two-dimensional counterparts: the column density PDFs (*11-12*) and the column density threshold of star formation (*13-14*). We must, however, accept that these cannot be used in the theories based on Eq. (1), because of the non-trivial transformation between the volume and column densities (*15-16*). An analytic technique to estimate $\rho$-PDFs from column densities exists (*16*), but is not widely applied because of its stringent requirements for the isotropy of the data. A technique exploiting molecular line observations also exists (*17*), but it samples the $\rho$-PDF sparsely, hampering the determination of its shape. To overcome the problem, some studies have derived

*SFRs* using the mean densities of the clouds instead (*18*). Even though reasonably successful in predicting *SFRs*, the approach does not connect the processes shaping the ISM to *SFRs* as directly as the theories utilizing Eq. (1). Consequently, exactly how those processes control star formation remains unknown.

To make progress, we develop an approach to estimate the $\rho$-PDF functions and the critical density from column density data (*19*). We represent the data as a set of hierarchical, three-dimensional (3D) structures. First, we decompose the column density maps with wavelet filtering to describe the structure at different spatial scales. Then, significant structures are identified at the different scales, and their 3D geometries are modeled with prolate spheroids. We chose this shape based on tests against numerical simulations (*19*). It allows modeling of both elongated, filament-like structures that are common in molecular clouds, and near-spherical shapes that represent small-scale, clumpy structures. The inclination angles of the spheroids are not known and are assumed to be zero. This leads to a high uncertainty in the densities of *individual* structures, but we show that when averaged over *numerous* structures, the $\rho$-PDF is reconstructed reasonably well (supplementary online text). The masses of the structures are calculated from the column densities at their respective scales. Finally, the hierarchical cloud structure is modeled by placing the overlapping structures inside each other's, allowing modeling of complicated, asymmetric structures. The volumes (d$V$) and masses, and hence densities (d$\rho$), of all structures are known, which yields the $\rho$-PDF.

We tested the technique with 14 numerical simulations of magneto-hydrodynamic, self-gravitating turbulence (*19*). The $\rho$-PDFs are reasonably well recovered under various physical conditions (Figs. S5-S10, *19*). The important $\rho$-PDFs parameters, the mean and width, have about 10% and 20% uncertainty, respectively (supplementary online text).

With this technique in hand, we derived $\rho$-PDFs for molecular clouds. As observational data, we employed column density maps derived from dust extinction mapping (*11*). We derived $\rho$-PDFs for 16 molecular clouds closer than 260 pc (Figs. 1, S1-S3). The derived $\rho$-PDFs probe the range of volume densities from 80 cm$^{-3}$ to 5×10$^4$ cm$^{-3}$. The sensitivity of our technique decreases above ~3×10$^4$ cm$^{-3}$, because the extinction maps cover a limited dynamic range of column densities (*19*). The $\rho$-PDFs closely follow log-normal functions, as predicted by turbulence theory (Eq. 2), and their widths vary between $\sigma_s$ = [1.2, 2.0] (Table S1).

Having quantified the $\rho$-PDFs, we can establish the relationship between the clouds' density structure and their star formation activity. As a measure of this activity, we adopted the number of young stellar objects, $N_{YSO}$, in the clouds (*19*). This number was used to estimate the mean star formation surface densities, $\Sigma_{SFR} = \dfrac{N_{YSO} <M>}{A \times 2 Myr}$, where $A$ is the cloud area, 2 Myr is the star formation time-scale (*13-14, 20*), and $<M>$ = 0.5M$_\odot$ is the mean stellar mass. We show that the $\rho$-PDF widths correlate with $\Sigma_{SFR}$ (Fig. 2A). This correlation invokes two possible interpretations. One is that the clouds' density structures evolve with time, characterized by the widening of their $\rho$-PDFs and consequent increase of $\Sigma_{SFR}$. Another interpretation is that the initial conditions of cloud formation set the clouds' density structures, which then control the $\Sigma_{SFR}$. Distinguishing between these scenarios with the available observational data is difficult (discussed in supplementary online text).

Once that we had quantified the $\rho$-PDFs and assessed their relation to star formation, we could estimate the critical density of star formation. Our sample includes three clouds on the verge of star formation; they have either formed only one star, or no stars at all. The mean of the highest log densities probed by the $\rho$-PDFs of these clouds is $s$ = 4.2 ± 0.3, which corresponds to a volume density of $(5 \pm 2) \times 10^3$ cm$^{-3}$ (*19*). This threshold does not depend strongly on the spatial resolution of the data we used (*19*). We interpreted these values as the critical densities in the clouds of our sample, noting that cloud-to-cloud variations may exist (*5*). Previously, the critical volume density based on analyses of observed column densities has been suggested to be ~$10^4$ cm$^{-3}$ (*13*) and $(6.1 \pm 4.4) \times 10^3$ cm$^{-3}$ (*21*) in nearby clouds. The observational estimates of the critical density are generally smaller than analytical model predictions that indicate $(2\text{-}5) \times 10^4$ cm$^{-3}$ (*5*). The reason for this discrepancy remains unknown.

The $\rho$-PDFs and critical density allow us to infer the *SFE* of star-forming gas. Following Eq. (1), only the gas above $s_{\text{crit}}$ participates in star formation. The mass of the star-forming gas is then: $M_{sf} = M(s > s_{crit}) = M_{cloud} \int_{s_{crit}}^{\infty} \frac{\rho}{\rho_0} p(s) ds$ (Table S1). The *SFE* of the gas above $s_{\text{crit}}$ (referred to as dense gas) is: $SFE(dense\ gas) = \frac{M_{stars}}{M_{stars} + M_{sf}}$, yielding the average of $16^{+20}_{-9}\%$ for our sample. The *SFE(dense gas)* is independent of the fraction of star-forming gas in the clouds (Fig 2B), indicating that it is independent of the density structure of the clouds' lower-density regions (supplementary online text).

The *SFE(dense gas)* we derive using volume densities is somewhat higher than an estimate based on column densities. We estimated the efficiencies also from the column densities, using $N=6.3 \times 10^{21}$ cm$^{-2}$ as the critical value (*13*). This yielded a mean efficiency of $6^{+11}_{-4}\%$. The masses of star-forming gas estimated from volume densities are lower than those estimated from column densities, yielding higher *SFE*. This difference results from the fact that when measured from column densities, $M_{sf}$ contains a contribution from a diffuse envelope that surrounds the dense (star-forming) gas. Our volume density technique removes this component.

In addition, when estimated from column densities, the *SFE* correlates with the dense gas mass fraction (Fig. 2B). This correlation is likely artificial; the relative contribution of the envelopes to the $M_\mathrm{sf}$ is larger when the dense gas fraction is lower.

The constraints we derived for the $\rho$-PDF, critical density, and *SFE* provide insight to the mass range of star-forming molecular clouds. Suppose the lowest possible mass of a star is given by the hydrogen-burning limit, 0.08 M$_\odot$. A simplistic calculation suggests that a minimum mass for a maternal molecular cloud to form a star is then of the order of 30 M$_\odot$ (*19*). This coincides with the mass range (5-50 M$_\odot$) of the smallest known star-forming molecular clouds, i.e., globules (*22*).

We similarly estimated the mass of a cloud likely to form high-mass (> 20 M$_\odot$) stars. The mass required for a natal molecular cloud to have a 95% probability to form a high-mass star is about $7^{+14}_{-5} \times 10^4$ M$_\odot$ (*19*). This is in agreement with the fact that from the Solar neighborhood clouds, only Orion A, whose mass is $10^5$ M$_\odot$ (*11*), is forming high-mass stars. For only a 50% probability to form a high-mass star, the same calculation yields a cloud mass of $1.5^{+3.4}_{-0.7} \times 10^4$ M$_\odot$ (*19*).

The $\rho$-PDFs can also help us to understand star formation on scales larger than individual molecular clouds. Our sample represents almost all molecular gas closer than 260 pc. The total $\rho$-PDF of the sample (Fig. S4) indicates that about 2.5% of this gas is above the star formation threshold. If we hypothesize that the total $\rho$-PDF is close to the average $\rho$-PDF of the Milky Way gas, a *SFR* of about 3 M$_\odot$/yr follows for the Milky Way (*19*). This agrees with other *SFR* estimates for the Milky Way (*23-25*). It is difficult to estimate how close the $\rho$-PDF of our sample is to the Galactic mean; the Galactic $\rho$-PDF is not generally known. Some works suggest that the turbulence properties are universal (*26*) and that the mean surface densities of gas do not depend strongly on the Galactic environment (*27*). This implies that the $\rho$-PDF of molecular gas may not vary much in general. Some observations have indicated that the $\rho$-PDFs of massive clouds may significantly differ from the local clouds (*28*). However, it is not known how representative such clouds are, or what is the relationship between density and star formation in them. In conclusion, albeit simplistic, our estimate of the Milky Way *SFR* lays out the possibility that the *SFR*s of entire galaxies are imprinted on their $\rho$-PDFs when averaged over hundreds-of-parsec scales.

With our approach to estimate molecular cloud $\rho$-PDFs, we were able to derive fundamental quantities that characterize star formation. The results provide an observationally established basis for predicting *SFR*s of molecular clouds, and they may also lead to a better understanding of the *SFR*s of entire galaxies.

**Acknowledgments:** We thank H. Beuther, P. Clark, and A. Stutz for discussions. J. K. acknowledges the Deutsche Forschungsgemeinschaft priority program 1573 ("ISM-SPP"), and C. F. the Discovery Projects Fellowship from the Australian Research Council (grant DP110102191). We acknowledge supercomputing time at Leibniz-Rechenzentrum (grant pr32lo), Jülich Supercomputing Centre (grant hhd20), and Partnership for Advanced Computing in Europe (grant pr89mu). This work is based on publicly available data from the 2-Micron-All-Sky-Survey (http://irsa.ipac.caltech.edu/Missions/2mass.html).

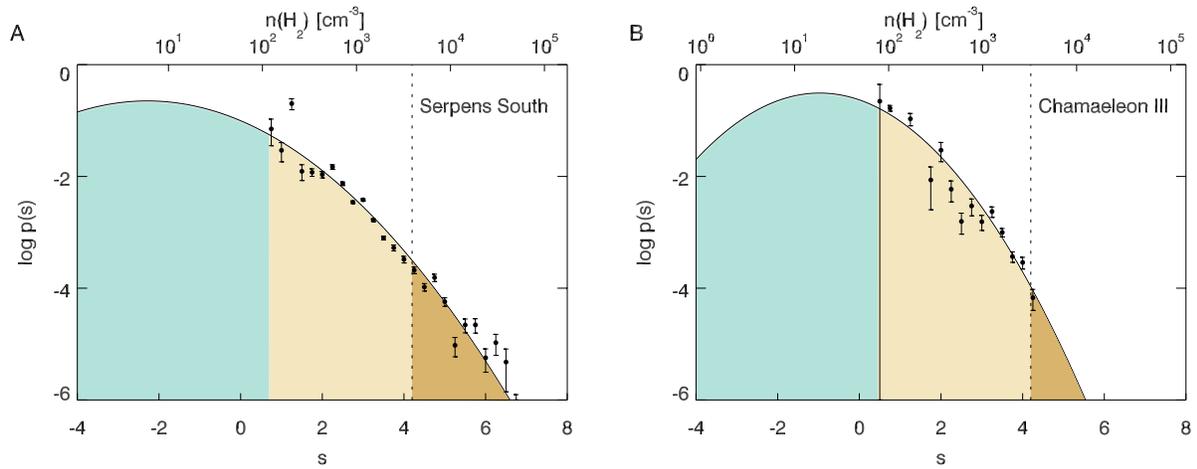

**Fig. 1**. $\rho$-PDFs of two molecular clouds. **A**, The star-forming Serpens South cloud. **B**, The non-star-forming Chamaeleon III cloud. The solid lines show fits of lognormal models. The dark brown color indicates the star-forming gas. The light brown color indicates the significant structures enveloping star-forming gas. The green color indicates the relatively non-structured gas.

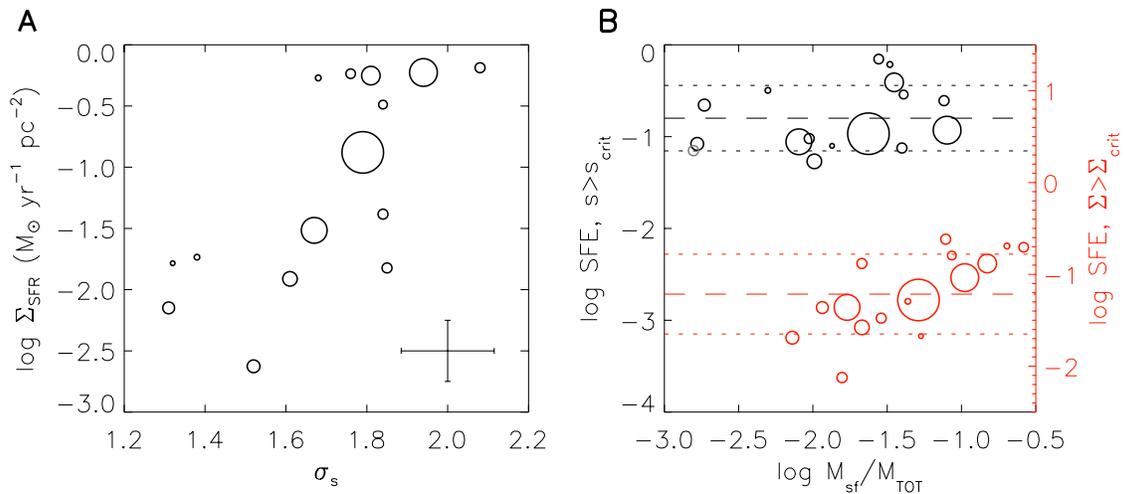

**Fig. 2.** Relationship between the star formation and molecular cloud density structure. **A**, *SFR* surface densities ($\Sigma_{SFR}$) as a function of the $\rho$-PDF width ($\sigma_s$). The symbol sizes reflect the cloud masses, spanning the range $0.07 - 3.6 \times 10^4$ $M_\odot$. The error bar indicates the 1-$\sigma$ statistical uncertainty. **B**, *SFE* of the star-forming gas as a function of the dense gas mass fraction (black circles, the left y-axis). The *SFE*s of the gas above the critical column density of $N=6.3 \times 10^4$ cm$^{-2}$ are also shown (blue circles, the right y-axis). The dashed line shows the mean and dotted lines the standard deviation.

**Supplementary Materials:**
Materials and Methods
Supplementary Text
Figs. S1-S9
References (29-57)
Table S1

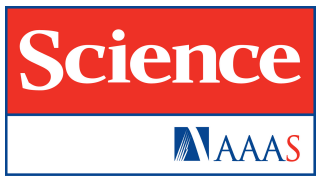

# Supplementary Materials for

## Unfolding the Laws of Star Formation: The Density Distribution of Molecular Clouds

Jouni Kainulainen, Christoph Federrath, Thomas Henning
correspondence to:  jtkainul@mpia.de

**This PDF file includes:**

    Materials and Methods
    SupplementaryText
    Figs. S1 to S9
    Table S1

**Other Supplementary Materials for this manuscript includes the following:**

**Materials and Methods**

Wavelet decomposition

The derivation of the $\rho$-PDFs from column density data was based on a spatial filtering algorithm known as the *à Trous* wavelet transform (*29*). We implemented the algorithm through a series of consecutive image convolutions and subtractions. We first decomposed the original column density maps into a series of scale-maps, $X_i$, that represent the structure of the data at various spatial scales. A series of high-pass filtered images was generated from the original image, $X$, via subtraction $X_i^{hp} = X - X_i'$, where $X_i'$ is the original image convolved with a Gaussian kernel that has a full-width at half-maximum (*FWHM*) width of $2^i$ resolution elements. Then, the scale-maps were created via subtraction $X_i = X_i^{hp} - X_{i-1}^{hp}$ for the scales $i > 1$. The scale-map $X_1$ equals to the high-pass filtered map $X_1^{hp}$. The number of scale-maps, $N_{scales}$, for each original map was decided by the condition $2^i < \min\{l_x, l_y\} / 2$, where $l_x$ and $l_y$ are the number of resolution elements in the map in $x$ and $y$ directions.

Structure identification

The scale-maps resulting from the wavelet decomposition were used to identify structures and estimate their masses and volumes. First, significant structures were identified from each scale-map using a threshold algorithm (*30*). The algorithm requires as an input the contour levels that are used in identifying structures. The lowest contour level marks the detection limit and was set to three times the standard deviation that was measured from the lowest-column density regions in the scale-maps (at each scale separately). These regions were chosen subjectively by eye, and they were typically about 30% in size with respect to the entire map. The contour spacing was also set to three times the standard deviation. The object identification was run for each scale-map. Only those structures at each scale $i$ were accepted as significant structures that overlapped with significant structures at scales $i$-1 and $i$+1. For two structures to overlap, it was required that the center point of a structure at scale $i$ coincides with any pixel of any structure at scales $i$-1 and $i$+1. In the case $i = 1$, only an overlap with a structure at scale $i$+1 was required. Similarly, in the case $i = N_{scales}$, only an overlap with a structure at scale $i$-1 was required. This requirement decreases considerably the number of detections that originate from an overlap of structures along the line of sight that generates local column density peaks, but which do not correspond to local volume density peaks (*31*).

3-dimensional structure modeling

The structure identification results in a set of structures at each spatial scale, i.e., for each scale-map. The volumes of these structures were estimated using a prolate spheroidal model (ellipsoid in two dimensions) that is in the plane of the observations. The threshold algorithm results in an estimate of the *FWHM* of the structure in two perpendicular image dimensions, and the total area covered by the structure, $A$ (*30*). The semi-major axis of the prolate spheroid was approximated as $a = \sqrt{A/f\pi}$, where $f = \min\{FWHM_x, FWHM_y\}/\max\{FWHM_x, FWHM_y\}$ is the aspect ratio. Then, the semi-minor axis is $b = fa$. and for a prolate spheroid in the plane of the sky $c = b$. The volume of the structure was then computed as $V = 4/3 \pi a^2 b$. Note that the volumes of individual structures are estimated only with a rather high uncertainty, because the real inclination angle of the spheroid is not known. Therefore, even though the final $\rho$-PDF that is composed of up to hundreds of structures is recovered reasonably well (see the *Method testing* below, Figs. S5-S8), the density structure of any *individual* structure is clearly less constrained. This is especially likely in the case of structures with extreme axis ratios such as long filaments. We also tested the technique by using spheres instead of prolate spheroids as 3-dimensional models. This approach produced strongly erroneous $\rho$-PDFs, resulting from the fact that the volumes of elongated structures that are common in both simulations and real clouds are not estimated well with spheres.

Hierarchical volume modeling
The structures at different scales that overlap in 2-dimensional projection are obviously very likely to overlap in 3-dimensional space. Therefore, it is necessary to estimate what fraction of the volume of each structure in scale $i$ is already assigned to structures at scales $1…i-1$, if any. Starting from the scale $i = 2$, we subtracted from the volume of each structure at the scale $i$ the volume(s) of the overlapping structure(s) at the scale $i$-1. If the structures overlapped only partially, then only the fraction that overlaps with the parental structure was subtracted. Typically, about half of the volume of a structure at scale $i$ was occupied by structures at scales $1… i$-1.

Masses of the structures
The masses of the structures were derived from their corresponding scale-maps, $X_i$. The signal (visual extinction) in the maps was transformed into the units of hydrogen column density using the conversion $N(H) = 1.9 \times 10^{21}$ cm$^{-2}$ $A_V$ (*32*). Then, the total mass of the structure was computed by summing up the column density in all pixels assigned to the structure, $M = N(H) \times A \times m_p \times 1.4$, where $A$ is the total area, $m_p$ is the proton mass and the factor 1.4 results from the addition of helium. The masses of the structures at the scales $1…i$-1 were subtracted from the masses of the structures at scale $i$.

$\rho$-PDF derivation
The structure modeling procedure results in an assemblage of structures whose masses, $M$, and mean volumes, $V$, are known. The division of these values gives the mean densities, $\rho$, of the structures. The $\rho$-PDF describes the fraction of volume, d$V$, in density d$\rho$. Thus, the volume fraction of gas in the given density immediately yields the $\rho$-PDF. In order to estimate the PDF of mean-normalized densities, a mean density $\rho_0$ needs to be estimated. This was calculated from the maps via $\rho_0 = M_{TOT} / Ad$, where $M_{TOT}$ is the total mass in the observed area, $A$ the total area of the map, and $d = \min\{l_x, l_y\}$, where $l_x$ and $l_y$ are the dimensions of the map in $x$ and $y$ directions. The area for the calculation was set so that it *1)* encircles the $A_V = 1$ mag contour, and *2)* is the smallest possible rectangle that does this. We note that this procedure could not be accomplished in the case of the Pipe Nebula and Lupus III, because of the diffuse Galactic dust component that confuses with the field at low Galactic latitudes. In these cases, we chose the field more subjectively with arbitrary placement. The final $\rho$-PDFs were then calculated from the logarithms of the densities that were normalized by the mean density calculated for these areas.

Method testing

We tested the technique by applying it to a set of numerical turbulence simulations that include self-gravity and sink particles to model star formation (*33-34*). We included in the test all those simulations whose resolution was $512^3$ computational cells. Each simulation consists of several time-steps. The first time-step corresponds to a state in which the turbulence is fully developed and the gravitation is switched on. Then, the evolution of the simulation under the influence of gravity is followed. The time since the onset of gravity is recorded as a function of the fractional mass accreted into the sink particles, i.e., of star formation efficiency (*SFE*). The simulations continue until *SFE* = 20%. We show the $\rho$-PDFs derived for 14 simulations at three time-steps, namely $t = 0$ (the initial, purely turbulent state), *SFE* = 10%, and *SFE* = 20% (Figs. S5-S8). We also discuss the correspondence between the derived $\rho$-PDFs and the true underlying $\rho$-PDFs (*supplementary online text*).

The molecular cloud sample
We derived $\rho$-PDFs for a sample of nearby molecular clouds at distances less than $d <$ 260 pc (*11*). From the 23 clouds in the sample of (*11*) we discarded "Coalsack", because its extinction map is hampered by a Galactic spiral arm cloud at a farther distance (*35*). We also could not derive a $\rho$-PDF for the "Lupus V" cloud, because of the small number of significant structures in it. We also did not include in the sample the five clouds at the distances between 260-500 pc, because the column density maps of those clouds were derived in different resolution compared to the more nearby clouds. Thus, our sample consists of 16 molecular clouds.

Dust extinction mapping
The column density data we employed were derived using a dust extinction mapping technique (*11*). The physical spatial resolution of the data is 0.1 pc, which approximately corresponds to the sonic length in the clouds. Thus, our data just reaches the scale that is used in some of the analytic star formation theories as a basis to infer the star formation threshold (*2-3*). In this resolution, the data probe column densities up to $N(H_2) \approx 30 \times 10^{21}$ cm$^{-2}$, depending on the Galactic coordinates of the cloud. For example, in Ophiuchus ($[l, b] \approx [354, 16]$) column densities up to $N(H_2) \approx 35 \times 10^{21}$ cm$^{-2}$ can be measured, while in Taurus ($[l, b] \approx 171, -16]$), column densities up to $N(H_2) \approx 25 \times 10^{21}$ cm$^{-2}$ can be measured. These upper limits also limit the volume density range that can be probed by the technique. It is not straightforward to estimate this upper limit, because it depends on the 3-dimensional structure of the cloud at the scale smaller than our resolution, which is unknown. We can perform a rough estimate of the upper limit by considering a maximum column density, say $N(H_2) \approx 30 \times 10^{21}$ cm$^{-2}$ inside an area (and volume) of one pixel (pixel size is half the physical resolution). This yields the maximum number density of $n(H_2) \approx 3 \times 10^4$ cm$^{-3}$. Note that since there is no direct mapping between column and volume densities, structures at volume densities higher than this can be detected. However starting from about this volume density, the derived $\rho$-PDFs may systematically underestimate the PDF. Only very few structures in all our clouds reach such high densities (see Figs. S1-S3).

$\rho$-PDF fitting

The ρ-PDFs derived from the observational data were fitted using lognormal functions. In addition to the data points derived using our approach, the fit is constrained by the normalization of the PDF, requiring $\int_0^\infty p(s)ds = 1$. This was used in the fits as a fixed boundary condition.

Young stars in the clouds
We correlated the derived ρ-PDF parameters with star formation activities of the clouds, using as a basis the number of identified young stellar objects (YSOs) in the clouds from the literature (*13, 36-40*). The details of the observational techniques with which the YSOs were identified vary from cloud to cloud, but they are always based on near-infrared, and possibly also mid-infrared, colors of the sources. Consequently, the completeness of the YSO number estimates is not uniform among the clouds. While the current observational data does not allow for a more uniform definition of the YSO populations, the YSO populations are likely correct within a factor two (*13*). The Chamaeleon III cloud has no identified YSOs. We used a value $N_{YSO} = 0.5$ to include it in Fig. 2B.

Star formation threshold
We derive the star formation threshold by considering the highest volume densities (in units of *s*) measured in the three clouds of our sample that have one or less identified YSOs in them (LDN1719, Musca, Cha III). The mean of the highest *s* values in these clouds is 4.2 ± 0.3 (Table S1). The mean of the volume densities corresponding to the highest *s* values is $(5 \pm 2) \times 10^3$ cm$^{-3}$.

We explored how the resolution of the extinction maps affects the star formation threshold. We derived a higher-resolution extinction map for the Musca cloud using deeper near-infrared data that yielded an extinction map for the cloud in the resolution of 0.03 pc (45"). These data were collected with the Very Large Telescope/HAWK-I instrument under the program ID 087.C-0822, PI J. Kainulainen. The critical log density value, $s_{crit}$, inferred for the cloud from those data was 0.1 higher than from the data in the resolution of 0.1 pc. We note that from the three clouds we use to evaluate the star formation threshold, Musca appears to be closest to active star formation. Consequently, it is likely that it also harbors the highest densities, and hence the effect of the resolution to the estimated critical density in it may be strongest. We conclude that the effect of resolution to the critical density ($s_{crit}$) is on the order of 0.1 and thus introduces an uncertainty of ~ 10%.

Minimum mass of a star-forming cloud
We made an order-of-magnitude estimate of the minimum mass of a cloud that can form stars using the ρ-PDFs of those clouds that are forming only a small number of stars and the star formation threshold. The mass of the least-massive stars is about the hydrogen-burning limit, i.e., 0.08 M$_\odot$. We estimate the mass of an initial dense core required to form such a star to be about 0.08 M$_\odot$ / 0.5 = 0.16 M$_\odot$, assuming the core-to-star efficiency $\varepsilon_{core} = 0.5$. While the formation process of a star in a real cloud undoubtedly depends on the physical characteristics prevalent in that cloud, such a constant efficiency appears to describe the average outcome of the process (*33, 41-42*). This mass of 0.16 M$_\odot$ needs to be above the density threshold $s > s_{crit} = 4.2$ determined using the $\rho$-PDFs. Adopting, as an example, the $\rho$-PDF parameters of the Musca cloud (Table S1) yields that about 1% of the cloud's mass is above $s_{crit}$. Thus, the minimum mass for an initial, maternal molecular cloud to form a dense core that, in turn, can form a star is $M = 0.16$ M$_\odot$ / 0.01 ~ 16 M$_\odot$. An alternative estimate can be calculated using the star formation efficiency value we derive, i.e., 16%. In this case the minimum mass of $M = 0.08$ M$_\odot$ / 0.16 / 0.01 ~ 50 M$_\odot$ follows. We give the mean of these two estimates, ~30 M$_\odot$ as an estimate of the minimum mass of a star-forming cloud. While these estimates span a wide range, they coincide with the mass-range of the Bok globules that are the lowest-mass star-forming clouds that we know of (*22*).

Mass of a cloud that can form O-type stars
We estimated the minimum mass of a cloud that is likely to form O-type stars ($M > 20$ M$_\odot$). About 4500 stars need to be randomly drawn from the initial mass function of stars (*43*) to have a probability of 95% for having the most massive star over $M > 20$ M$_\odot$ (*44*). Adopting 0.5 M$_\odot$ for the mean mass of stars and 16% star formation efficiency in the gas above $s_{crit}$, about 0.5 M$_\odot$ × 4500 / 0.16 ≈ 14000 M$_\odot$ of gas is needed above $s_{crit}$ to form (with 95% probability) one O-type star. Extrapolating from the most active clouds in our sample (Fig. 2A, Table S1), a cloud with 4500 members has the $\rho$-PDF parameters approximately $\mu = -2.5$, $\sigma_s = 2.2$. Integrating over the $\rho$-PDF with these parameters results in about 21% of mass being above $s_{crit}$. Thus, the minimum mass of a maternal molecular cloud that can (statistically) form O-type stars is 14000 M$_\odot$ / 0.21 ~ $7^{+14}_{-5} \times 10^4$ M$_\odot$. The quoted uncertainty results from the uncertainty of the assumed efficiency ($16^{+20}_{-9}\%$). The same calculation requiring the probability of 50% to have a high-mass star results in the cloud masses of about $1.5^{+3.4}_{-0.7} \times 10^4$ M$_\odot$.

Star formation rate of the Milky Way
We use the $\rho$-PDF compiled of all 16 molecular clouds to make a rough estimate of the Milky Way star formation rate. The total $\rho$-PDF has the parameters $\mu = -2.01$, $\sigma_s = 1.72$. Given these parameters, about 2.5% of the gas is above $s_{crit}$. The total molecular gas mass in the Milky Way is about $1.6 \times 10^9$ M$_\odot$ (*45-46*), and hence about $3.7 \times 10^7$ M$_\odot$ is above $s_{crit}$. If this gas forms stars with 16% efficiency in a time-scale of 2 Myr (*13*), a total star formation rate of the Milky Way of $SFR \approx 3$ M$_\odot$/yr follows. In this calculation, we assume that the *average* $\rho$-PDF of the Milky Way is close to that of our sample clouds (not that the gas $\rho$-PDFs is *the same* everywhere in the Milky Way). This is a strong assumption; there is no knowledge available on the $\rho$-PDFs in the context of the entire Milky Way. Our sample consists only of relatively low-mass clouds ($M < 10^5$ M$_\odot$) and does not include any giant molecular clouds that dominate the star formation in the Milky Way. Extensive works employing CO line emission observations have established that the characteristic properties of turbulence are similar over large range of environments in the Milky Way (*26*). However, the capability of CO measurements to efficiently capture variance in turbulence properties can be questioned, because of the limited dynamic range of densities the CO measurements probe. Recent works also suggest that the $\rho$-

PDFs of massive infrared dark clouds that are likely located in the Spiral arm regions are different from the Solar neighborhood clouds (*28*). However, no detailed information exists on the star-formation properties of those clouds, and therefore it is not possible to assess the relation between the star formation and density structure in them. It is also not known how common such massive clouds are. In summary, we find the predicted Milky Way SFR a reasonable scale estimate, but note that the grounding assumption behind the estimate cannot be tested with the current data.

**Supplementary Text**

Models used for the $\rho$-PDF function

We analyzed the $\rho$-PDFs of molecular clouds under the framework of supersonic turbulence that predicts approximately lognormal $\rho$-PDFs for isothermal, non-gravitating, non-magnetized media. This shape is in good agreement with our observations. In this framework, the $\rho$-PDF width depends on the physical conditions in the clouds via

$$\sigma_s^2 = \ln(1 + b^2 M_s^2 \frac{\beta}{\beta+1})$$

, where $b$ is the forcing parameter, $M_s$ the sonic Mach number, and $b$ the ratio of thermal to magnetic pressures. Recent studies have shown that the Solar neighborhood clouds are consistent with $b \approx 0.33$ (*28, 47*) and $M_s$ = 7-14 (*28*). These parameters yield the range of predicted widths of $\sigma_s$ = 0.8-1.8, where the lower limit corresponds to the non-magnetic case ($\beta \gg 1$) and the upper limit to a moderately magnetized case ($\beta$ = 0.2). The latter of these is likely more realistic, because in the current view molecular clouds are believed to have significant magnetic fields (*48*). The forcing parameter $b$ may also vary between the clouds, contributing to the observed width range (*47*). Thus, the variation of the observed $\rho$-PDF widths can potentially be explained by variations in the physical properties of turbulent media.

However, most of the clouds in our sample are actively forming stars, and the $\rho$-PDFs may well be affected by gravity in addition to turbulent motions (*33, 49, 50*). Strongly self-gravitating systems are predicted to develop an exponential $\rho$-PDF, $p(s) = e^{-3s/\kappa}$, where $\kappa$ is the exponent of the radial density distribution, $\rho \propto r^{-\kappa}$, that in the case of a collapsing cloud equals to $\kappa$ = 2 (*51*), which provides an upper limit for the feasible values of $\kappa$. Since the clouds of our sample have different levels of star formation activity, it can be hypothesized that self-gravity affects their density structure at different levels. To place the derived $\rho$-PDFs into this context, we fitted their high-density parts also with exponential functions. For most clouds, the exponential function provides an equally good fit to the high-density parts as the lognormal function (Figs. S1-S3), and we find a range of exponents $\kappa$ = [1.2, 2.0] (Table S1). However, the combined $\rho$-PDF of all clouds is better fit by a lognormal function (Fig. S4). The statistical uncertainty of the data points in the combined $\rho$-PDF is much lower than that of individual clouds. This suggests that the underlying $\rho$-PDF shape is lognormal instead of an exponential. The most active star-forming clouds show exponents close to $\kappa \approx 2$, while the exponents of the clouds with less star formation are smaller, $\kappa \approx 1.2$. This correlation is in agreement with a hypothesis that the $\rho$-PDFs of the clouds that are more active in star formation may be affected more by gravity than those of the clouds that form fewer stars. However, the correlation does not uniquely prove this hypothesis (discussed more below). This relationship between the star-forming status and density structure has previously been detected in the *column* density PDFs of molecular clouds (*11, 12, 52*); our work now recovers it also from the *volume* density structure of molecular clouds.

The two models discussed above represent two extreme models for $\rho$-PDFs: the former (log-normal) is a result of ideal, isothermal turbulence, and the latter a result of dominating gravity. In real clouds, both of these processes play a role, and quite possibly, their relative significance changes during the cloud evolution. This instigates a need for a theoretical framework that would encapsulate the effects of both processes. Such a framework is not currently available, but theoretical works are evolving in this direction (*53*).

The connection between the $\rho$-PDFs and star formation activity
We showed that there is a correlation between the star formation activity of molecular clouds and their $\rho$-PDF shapes (measured either by $\sigma_s$ or $\kappa$, see Fig. 2A). This correlation can be interpreted either as *i*) an evolutionary sequence in which young clouds have narrow $\rho$-PDFs, and as the star formation proceeds, their $\rho$-PDFs widen. For a given sample of clouds, this results in a range of $\rho$-PDF widths that reflects the range of evolutionary phases of the individual clouds in the sample; or *ii*) a static scenario in which the star-formation potential of a cloud is determined by its fiducial $\rho$-PDF that does not evolve as a function of time, or evolves significantly less than in the former scenario. As discussed earlier, variations in the gas properties can potentially induce the range of the observed $\rho$-PDF widths. It is not straightforward to distinguish between these two interpretations, because observations only represent temporal snapshots during the cloud evolution.

Statistical arguments can be used to explore which scenario might be prevalent. If the clouds' $\rho$-PDFs do not evolve significantly with time, clouds should exist whose $\rho$-PDF is wide ($\sigma_s \approx 2.0$ or $\kappa \approx 2$), but which have not (yet) formed any significant amount of stars. We do not have such clouds in our sample. Similarly, active star-forming clouds with relatively narrow $\rho$-PDFs should exist, but they are absent from our sample. However hampering these statistical arguments, the sample only includes 16 clouds and these clouds could be absent by a coincidence. This is especially possible if the time-scale of the starless phase during the molecular cloud evolution is short compared to the time-scale of the star-forming phase. Thus, by itself, the correlation between the observed $\rho$-PDF shapes and the cloud's star-forming activities does not imply an evolutionary sequence. Varying physical conditions in the clouds can also cause a range of $\rho$-PDF shapes. It is an interesting avenue for future studies with larger cloud samples to consider how much the physical conditions in the clouds reflect to their $\rho$-PDFs. Of especial interest in this respect are the virial parameters of the clouds, defined as the ratio of kinetic to potential energies, $\alpha = \dfrac{5\sigma_v^2 R}{GM}$, where $\sigma_v^2$ is the velocity dispersion, *G* the gravitational constant, and *R* and *M* the radius and mass of the cloud, respectively. Theoretical arguments suggest that *SFRs* anti-correlate strongly with virial parameters (*5*). Given our results (Fig. 2A), an anti-correlation between virial parameters and $\rho$-PDF widths is therefore expected.

Further insight into the origin of the $\rho$-PDF – star-forming activity correlation can be gained by setting them in the context of numerical and theoretical cloud evolution studies. Numerical simulations unanimously predict that the $\rho$-PDF of turbulent, self-gravitating media evolves with time (*33, 34, 49, 50, 54*), likely in time-scales comparable to the star-formation time-scales of molecular clouds. The strength of the $\rho$-PDF evolution in the simulations is coupled to the virial parameter in them. Following observational results (*55*), the virial parameter is typically set on the order of unity. The molecular clouds in our sample also show virial parameters of this order, $\alpha = [1, 7]$ (*11, 28*). When compared against these simulation studies, the range of the observed $\rho$-PDFs is most naturally understood as an evolutionary sequence, although, the data we use in this work does not allow establishing that directly. In this interpretation, the clouds' density structures are initially dominated by turbulent motions and characterized by a lognormal $\rho$-PDF that results from the predictions for turbulent media. As the clouds further evolve under the influence of gravity, their $\rho$-PDFs widen and develop toward the exponential shape predicted for strongly self-gravitating systems. Therefore, in the light of the current numerical simulations, the observations suggest an evolutionary scenario between the clouds' density structure and their star-forming activity.

If we hypothesize now that the observed $\rho$-PDFs form an evolutionary sequence, important consequences follow. First, only the $\rho$-PDF of relatively young clouds whose structure is not yet dominated by gravity carry the imprints of turbulence properties such as $\sigma_s - M_s/b/\beta$ relationship. When star formation is actively on-going, the $\rho$-PDF reflects rather the star-forming efficiency of the cloud than the initial turbulence properties. Second, the $\rho$-PDFs evolve over the entire range of molecular cloud densities, i.e., not only at high densities where the star formation factually takes place. The mass available for star formation, i.e., mass above $s_{crit}$, is expected to increase with time. Qualitatively, this is consistent with the picture of accelerating star formation in molecular clouds that results from slow gravitational contraction of a cloud (*56*). This increase of star forming gas should be taken into account when estimating the star formation rate based on Eq. (1). Also, in this interpretation, the evolution over the entire $\rho$-PDF shows that gravity has a significant impact on cloud structure at all densities, and hence all size-scales, present in molecular clouds, as shown for size-scales up to a few pc by earlier studies (*57*).

Accuracy of the $\rho$-PDF recovery in numerical simulations
We studied the feasibility of the $\rho$-PDF derivation technique by applying it to a set of 14 numerical simulations (*33, 34*). Figures S5-S8 show the $\rho$-PDFs derived for all simulations at three time-steps. Generally, the correspondence between the derived $\rho$-PDFs and the true underlying $\rho$-PDFs is good roughly above $s > 0.5$. At lower densities, the derived $\rho$-PDF becomes unreliable and commonly shows low values compared to the true $\rho$-PDF. This results from the fact that the mean column density of the real structures becomes close to the mean column density of the domain, and therefore disentangling them from random column density variations becomes uncertain. Importantly, the relative shapes of the $\rho$-PDFs are recovered well, regardless whether the underlying shapes are approximately lognormals (e.g., simulations No. 11, 12, 18, time-step *SFE*=0%) or power-laws (e.g., simulations No. 12, 14, 15, time-step *SFE*=20%). In one simulation (No. 24, as in ref. *33*), the $\rho$-PDF is severely over-estimated throughout the density range. However even in this case, the derived $\rho$-PDF reflects correctly the shape of the $\rho$-PDF. This simulation corresponds to the most extreme choice of the forcing parameter in the simulations, $b=1$ (the forcing parameter describes the ratio of the kinetic energy in compressive and solenoidal modes in the simulations; the value $b=1$ refers to a situation where all energy is in compressive modes). In general, the $\rho$-PDFs derived for simulations with the forcing parameter $b=1$ have larger scatter than those derived for simulations with lower values of $b$. This is magnified in the case of extremely strong gravity (simulation No. 5 with $\alpha = 0.07$). In this case the scatter in the data points of the derived $\rho$-PDF is large. However, molecular clouds of our observational sample do not reflect this kind of extreme environments.

The relation between the derived and true $\rho$-PDFs is further illustrated in Fig. S9 that shows the ratio of the derived $\rho$-PDF values to the true $\rho$-PDFs values as a function of the normalized density, $s$. The mean ratio is within 25% of the true value up to $s = 6$ and within 60% up to $s \approx 8$. Note that the $\rho$-PDFs derived from observations continue only up to $s = 5$-6. We further show that the first and second moments of the $\rho$-PDFs are recovered well (Fig. S9). The moments were calculated over the range of $s$ values for which the $\rho$-PDFs were derived, typically over $s=0$-6. The mean ratio of the derived and true first and second moments are $1.05 \pm 0.10$ and $0.99 \pm 0.20$, respectively. For two cases (out of the total of 42) the mean is overestimated erroneously, by a factor of about two. We also show in Fig. S9 the mean and standard deviation of the first and second moments computed from the $\rho$-PDFs we derive from extinction maps. We conclude that the uncertainties in the mean and width estimates are about 10% and 20%, respectively.

The good absolute correspondence between the derived and true $\rho$-PDFs is partly instigated by the fact that the numerical simulation domain has equal dimensions in all three spatial directions, i.e., it is box-shaped. Consequently, the line-of-sight depth of the domain, and hence its mean density, is always estimated correctly. In the case of observations this is not necessarily the case, and therefore, the depth of the cloud, and hence its mean density, may be misestimated by a relatively large factor. This can affect the scaling of the physical densities ($n$, or $\rho$) compared to the dimensionless densities ($s$, see Eq. 2), and from therein, the threshold density of star formation. It is not straightforward to quantify the (systematic) uncertainty arising from the uncertainty in the mean density. The aspect ratios of the clouds in our sample in two dimensions span the range ~0.5-1. Therefore, it seems likely that the uncertainty in the mean density is about 30%.

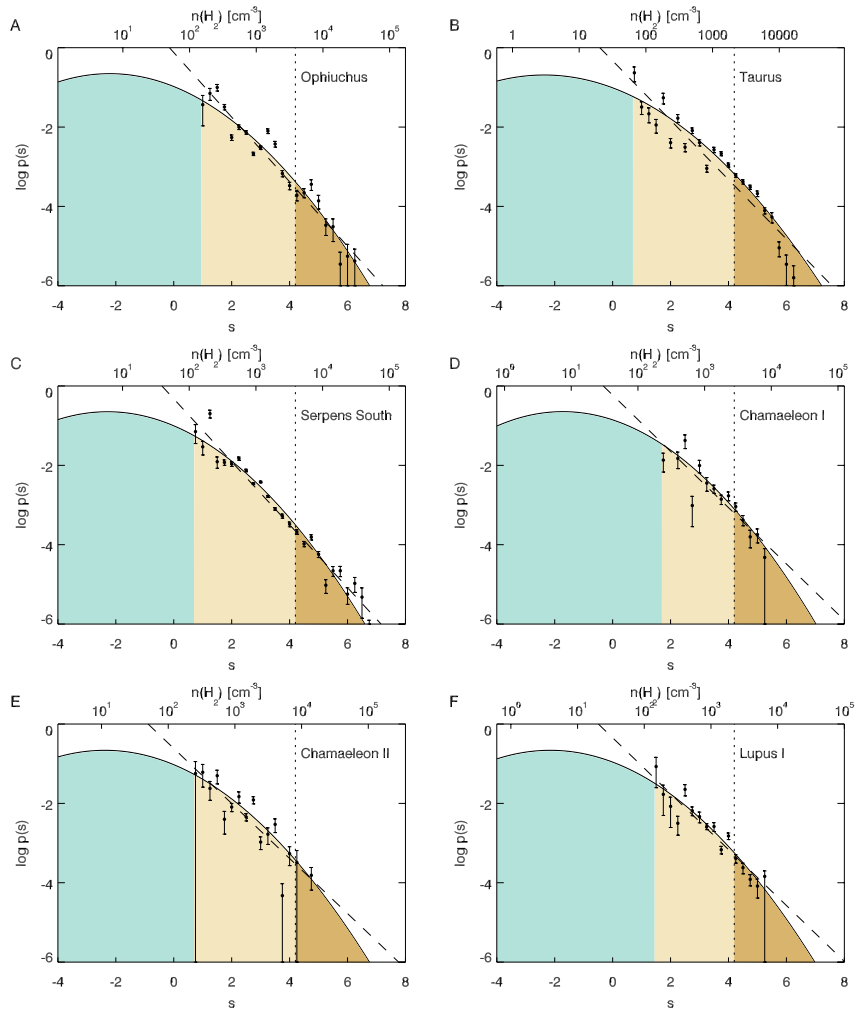

**Fig. S1.** $\rho$-PDFs of nearby molecular clouds. A. Ophiuchus; B. Taurus; C. Serpens South; D. Chamaeleon I; E. Chamaeleon II; F. Lupus I. In each panel, the solid line shows a fit of a lognormal function. The dashed line shows a fit of an exponential function. The parameters of the fits are given in Table S1. The vertical dotted line shows the density threshold of star formation $s_{crit} = 4.2$. The dark brown color refers to the star-forming gas in the cloud. The light brown color refers to the regime on which gas is organized into dense structures.

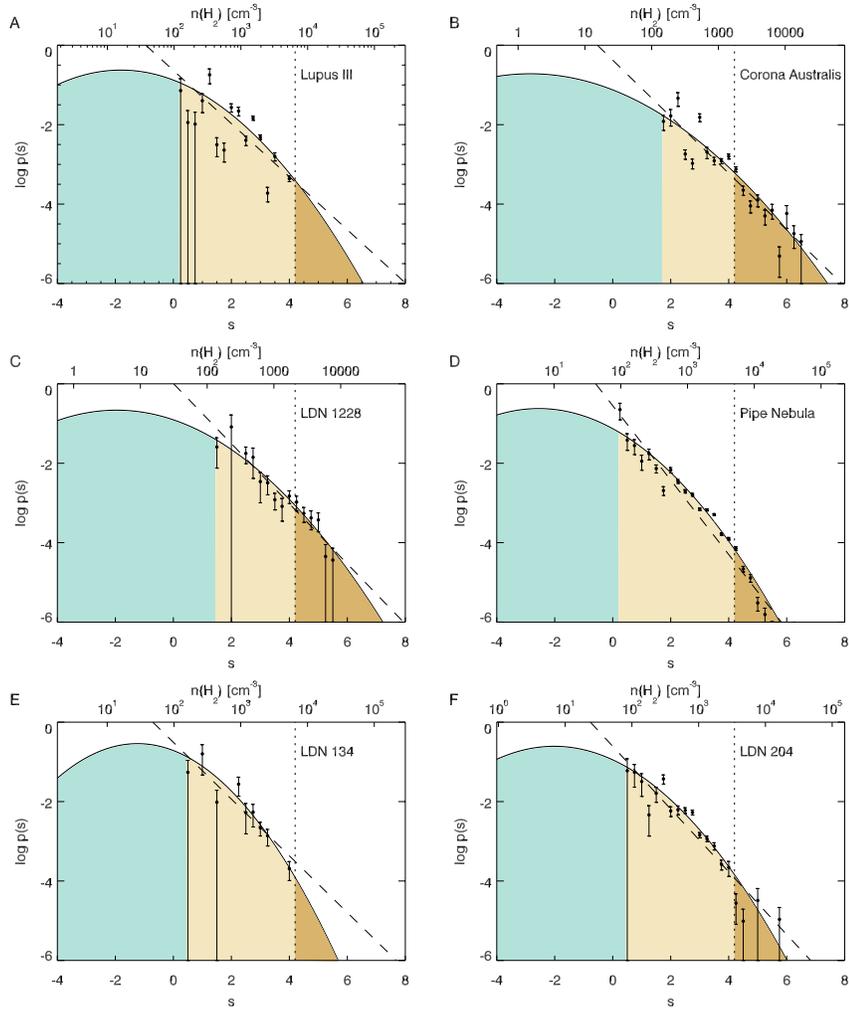

**Fig. S2.**

$\rho$-PDFs of nearby molecular clouds. A. Lupus III; B. Corona Australis; C. LDN 1228; D. The Pipe Nebula; E. LDN 134; F. LDN 204. In each panel, the solid line shows a fit of a lognormal function. The dashed line shows a fit of an exponential function. The parameters of the fits are given in Table S1. The vertical dotted line shows the density threshold of star formation $s_{\mathrm{crit}} = 4.2$. The dark brown color refers to the star-forming gas in the cloud. The light brown color refers to the regime on which gas is organized into dense structures.

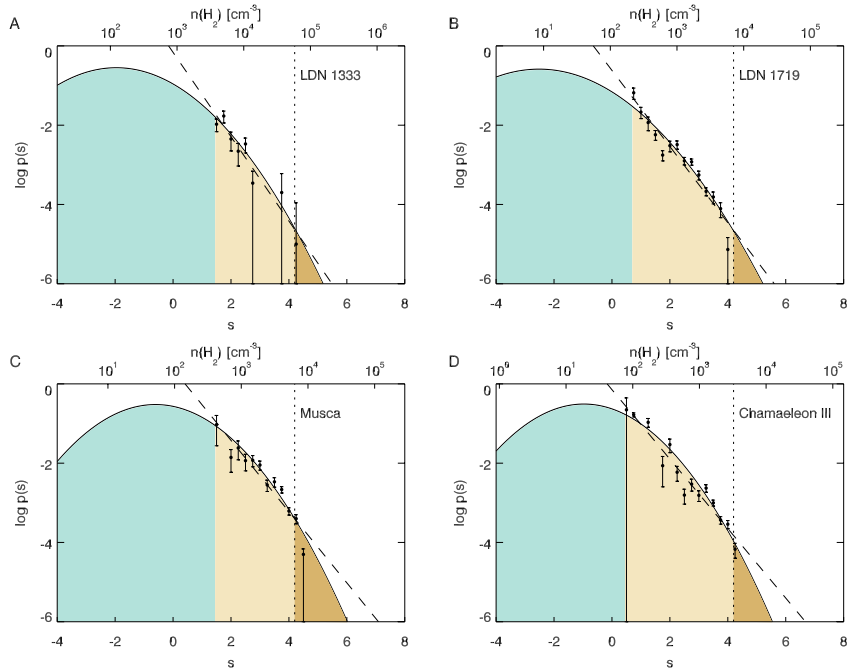

**Fig. S3**

$\rho$-PDFs of nearby molecular clouds. **A.** LDN 1333; **B.** LDN 1719; **C.** Musca; **D.** Chamaeleon III. In each panel, the solid line shows a fit of a lognormal function. The dashed line shows a fit of an exponential function. The parameters of the fits are given in Table S1. The vertical dotted line shows the density threshold of star formation $s_{\text{crit}} = 4.2$. The dark brown color refers to the star-forming gas in the cloud. The light brown color refers to the regime on which gas is organized into dense structures.

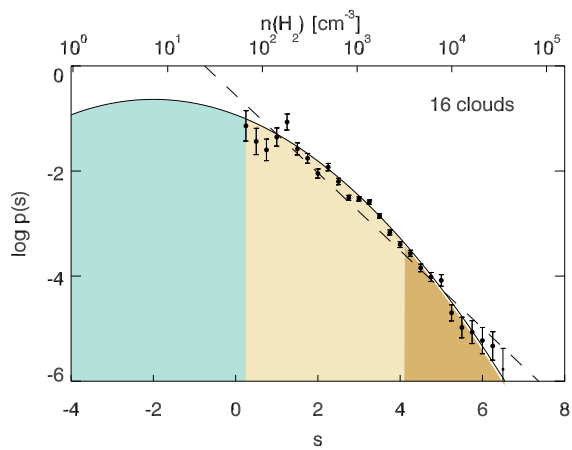

**Fig. S4**

The total $\rho$-PDF of the sample of 16 molecular clouds. The solid line shows a fit of a lognormal function. The dashed line shows a fit of an exponential function. The dark brown color refers to the star-forming gas in the cloud. The light brown color refers to the regime on which gas is organized into dense structures. The green color refers to the regime of diffuse, relatively non-structured gas. Using the total $\rho$-PDF, we derived the Milky Way star formation rate of 3 $M_\odot$ / yr.

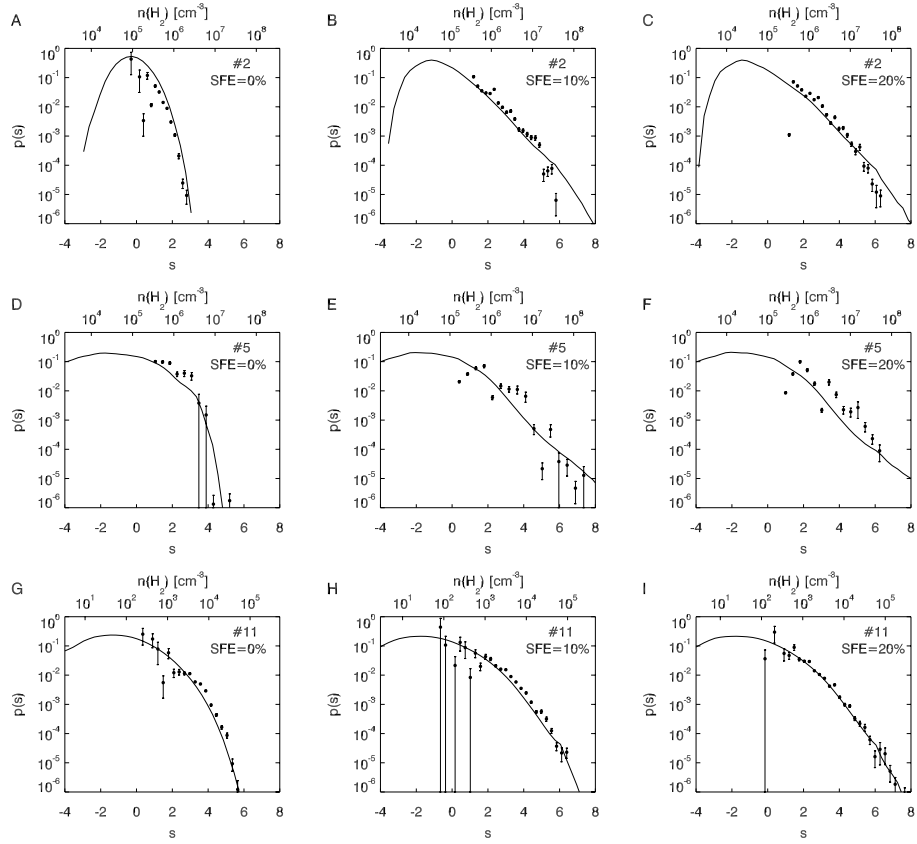

**Fig. S5**

ρ-PDFs derived for numerical simulations (*33*) (black data points). The solid lines show the true underlying ρ-PDFs that are composed of the 3-dimensional density data in the simulations. Each row shows the simulation at the time steps *t* = 0, *SFE* = 10%, and *SFE* = 20%. **A-C.** Simulation #2. **D-F.** Simulation #5. **G-I.** Simulation #11. The simulation number given in each panel corresponds to the number in Table 2 of ref. (*33*), providing a detailed list of simulation parameters.

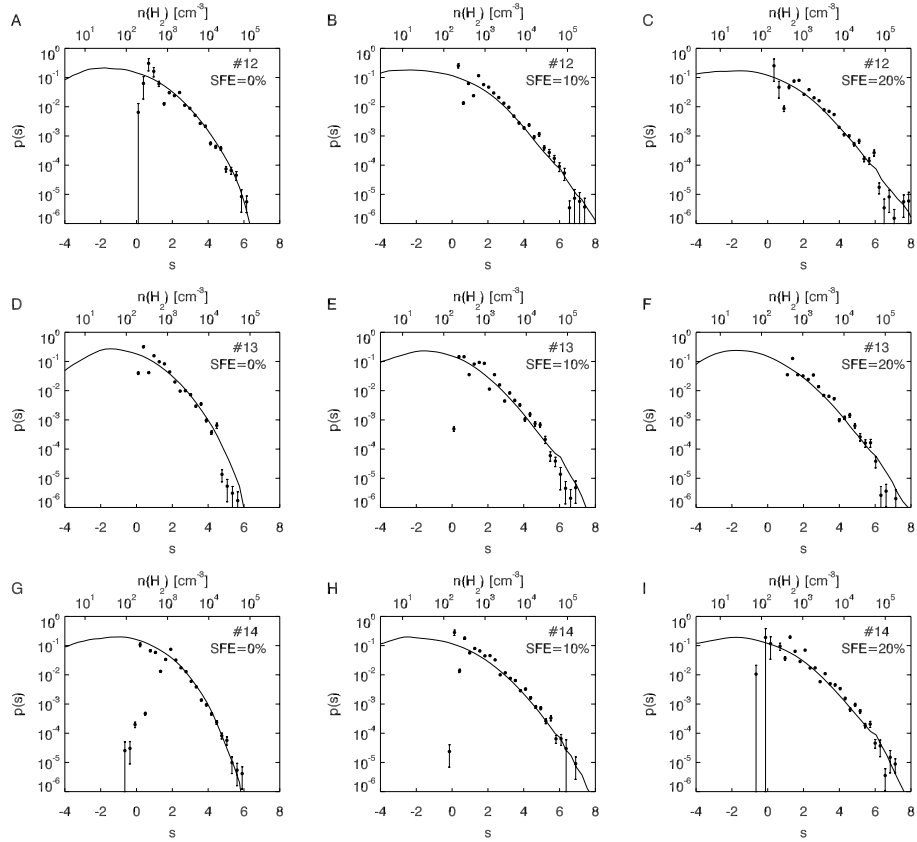

**Fig. S6**

$\rho$-PDFs derived for numerical simulations (*33*) (black data points). The solid lines show the true underlying $\rho$-PDFs that are composed of the 3-dimensional density data in the simulations. Each row shows the simulation at the time steps $t = 0$, *SFE* = 10%, and *SFE* = 20%. **A-C.** Simulation #12. **D-F.** Simulation #13. **G-I.** Simulation #14. The simulation number given in each panel corresponds to the number in Table 2 of ref. (*33*), providing a detailed list of simulation parameters.

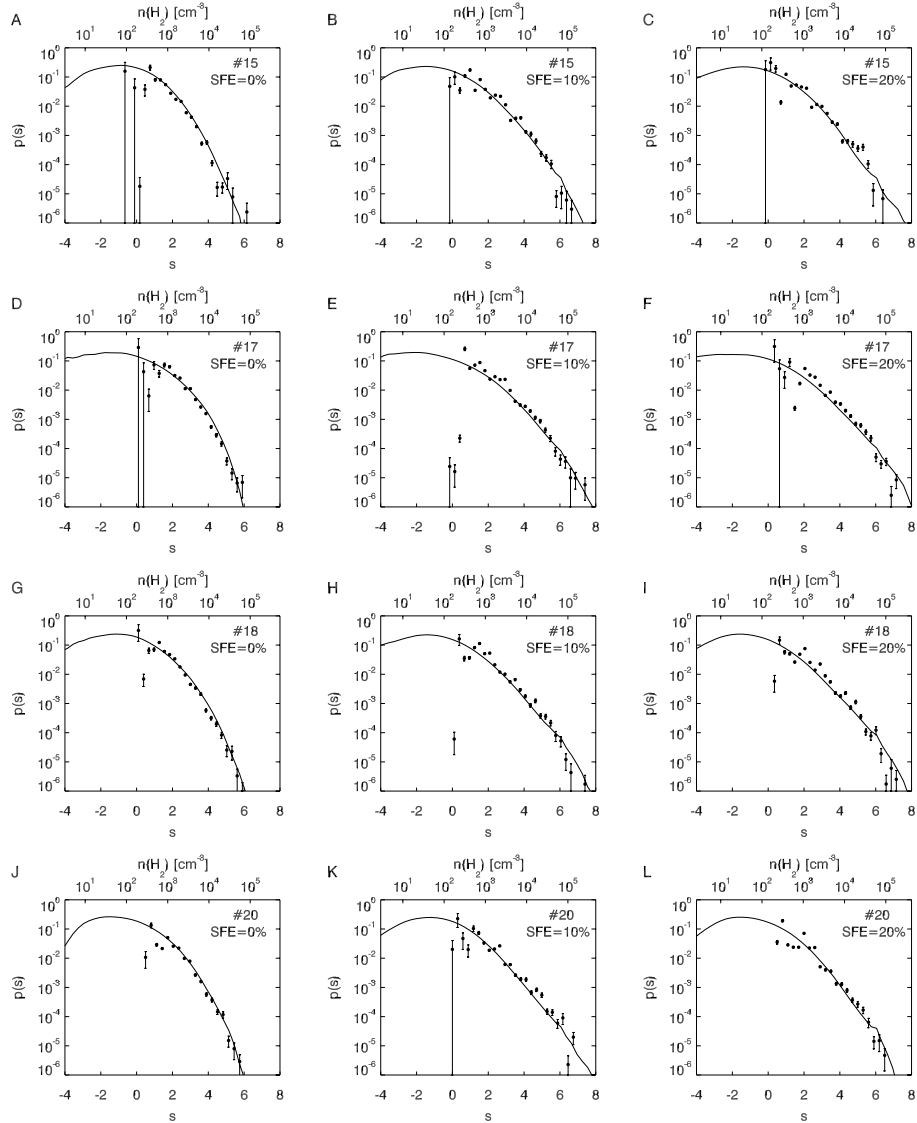

**Fig. S7**

$\rho$-PDFs derived for numerical simulations (*33*) (black data points). The solid lines show the true underlying $\rho$-PDFs that are composed of the 3-dimensional density data in the simulations. Each row shows the simulation at the time steps $t = 0$, *SFE* = 10%, and *SFE* = 20%. A-C. Simulation #15. D-F. Simulation #17. G-I. Simulation #18. J-L. Simulation #20. The simulation number given in each panel corresponds to the number in Table 2 of ref. (*33*), providing a detailed list of simulation parameters.

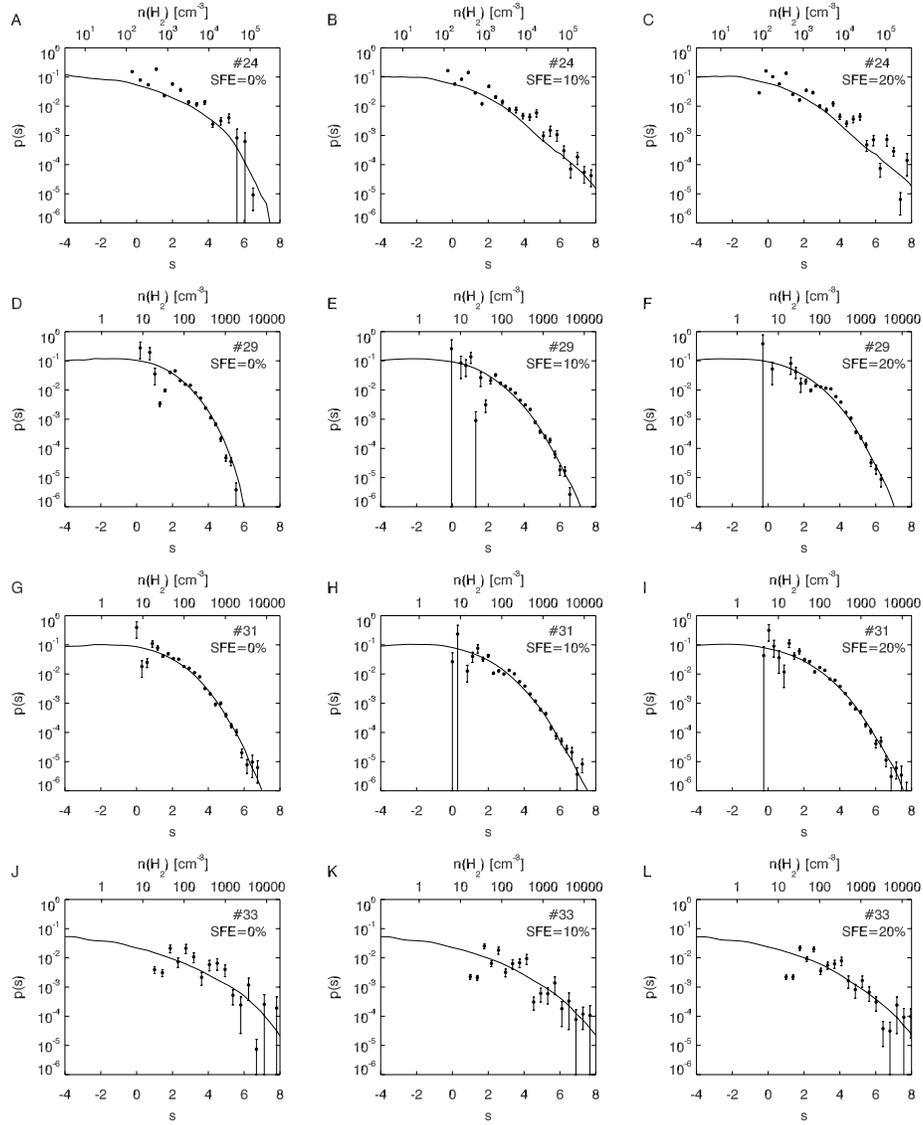

**Fig. S8**

$\rho$-PDFs derived for numerical simulations (33) (black data points). The solid lines show the true underlying $\rho$-PDFs that are composed of the 3-dimensional density data in the simulations. Each row shows the simulation at the time steps $t = 0$, *SFE* = 10%, and *SFE* = 20%. **A-C.** Simulation #24. **D-F.** Simulation #29. **G-I.** Simulation #31. **J-L**. Simulation #33. The simulation number given in each panel corresponds to the number in Table 2 of ref. (*33*), providing a detailed list of simulation parameters.

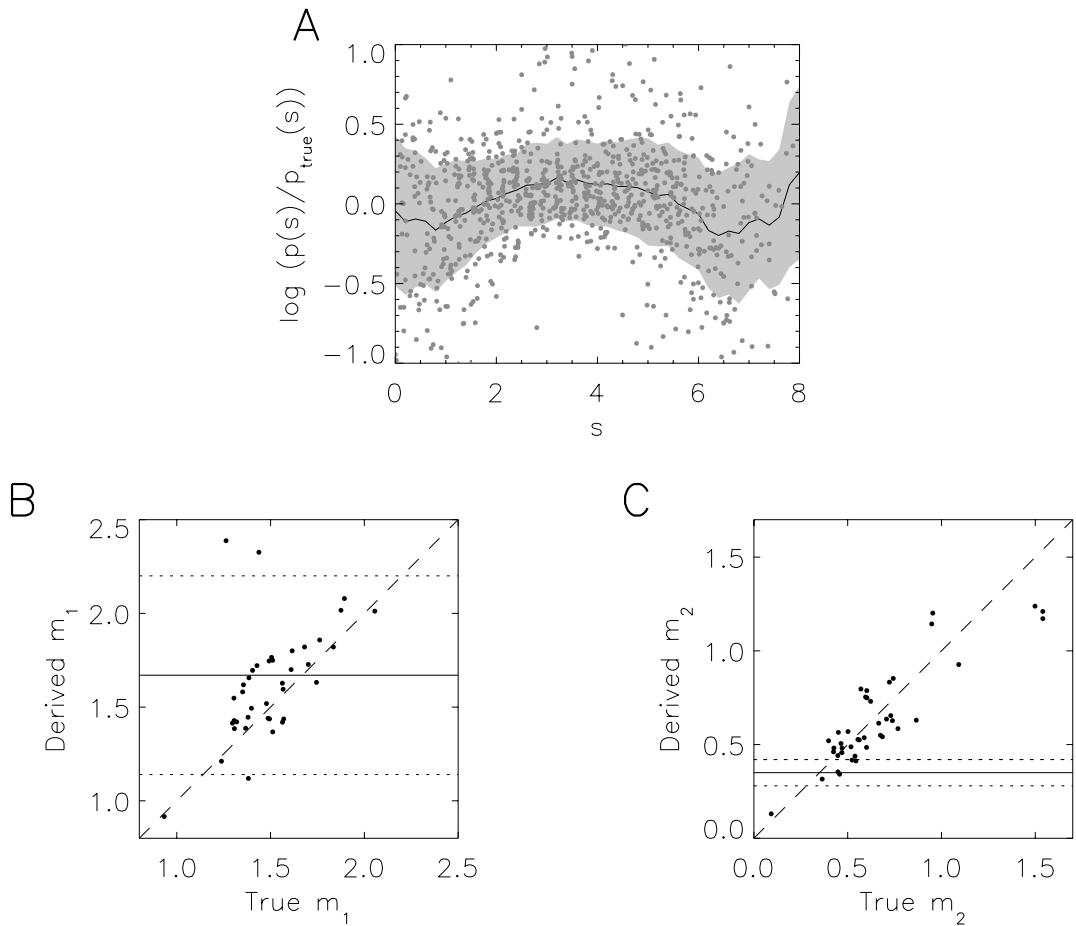

**Fig. S9**

Accuracy of the $\rho$-PDF derivation technique. **A:** The ratio of the derived $\rho$-PDFs to the true underlying $\rho$-PDFs in numerical simulations. The solid line shows the mean of the data points (grey circles) within a window that has a width of 0.4. The shaded grey area indicates one standard deviation. **B:** The relationship of the derived first moments ($m_1$), and **C:** second moments ($m_2$) to the true values. The dashed line indicates the one-to-one correlation. The horizontal solid and dashed lines indicate the mean and standard deviation calculated from the $\rho$-PDFs derived for the real molecular clouds.

**Table S1.**

Properties of molecular clouds.

| Cloud name | $\mu$ | $\sigma_s$ | $n_0$ (cm$^{-3}$) | max($s$) | $\kappa$ | $M_{sf}$ ($M_{sun}$) | $M_{sf}/M_{tot}$ (%) | $M_{Av>7mag}$ ($M_{sun}$) | $N_{YSO}$ |
|---|---|---|---|---|---|---|---|---|---|
| Ophiuchus | -2.21 | 1.81 | 115 | 6.2 | 1.95 | 247 | 4.1 | 1046 | 316 |
| Taurus | -2.36 | 1.94 | 64 | 6.1 | 1.71 | 1257 | 7.4 | 1653 | 335 |
| Serpens South | -2.29 | 1.79 | 117 | 6.7 | 1.58 | 834 | 3.3 | 1812 | 201 |
| Chamaeleon I | -1.72 | 1.76 | 210 | 5.2 | 2.05 | 51 | 5.4 | 482 | 237 |
| Chamaeleon II | -2.38 | 1.84 | 242 | 4.8 | 2.05 | 62 | 4.1 | 131 | 50 |
| Lupus I | -2.16 | 1.84 | 67 | 5.3 | 1.67 | 80 | 5.3 | 43 | 13 |
| Lupus III | -1.81 | 1.68 | 195 | 4.0 | 1.76 | 22 | 2.9 | 135 | 69 |
| Corona Australis | -2.84 | 2.08 | 52 | 7.8 | 1.83 | 153 | 9.6 | 157 | 100 |
| LDN 1228 | -1.96 | 1.85 | 62 | 5.3 | 1.53 | 19 | 6.9 | 58 | 4 |
| Pipe | -2.55 | 1.67 | 122 | 6.1 | 1.42 | 109 | 0.88 | 230 | 21 |
| LDN 134 | -1.24 | 1.38 | 196 | 4.0 | 1.96 | 3.2 | 0.51 | 28 | 3 |
| LDN 204 | -2.02 | 1.61 | 103 | 5.6 | 1.67 | 44 | 1.2 | 92 | 5 |
| LDN 1333 | -1.67 | 1.31 | 60 | 4.3 | 1.17 | 4.2 | 0.15 | 33 | 3 |
| LDN1719 | -2.44 | 1.52 | 157 | 4.0 | 1.39 | 5.5 | 0.22 | 24 | 1 |
| Musca | -0.61 | 1.32 | 170 | 4.5 | 1.54 | 5.8 | 1.0 | 23 | 1 |
| Chamaeleon III | -0.97 | 1.29 | 98 | 4.2 | 1.24 | 3.3 | 0.32 | 33 | 0 |
| The total PDF of 16 clouds | -2.01 | 1.70 | 102 | 6.8 | 1.61 | 2375 | 2.5 | 5980 | 1359 |